\documentclass{amsart}
\usepackage{amssymb,amsmath,amscd, amsthm,stmaryrd,verbatim, mathrsfs}

\newtheorem{Thm}{Theorem}
\newtheorem{theorem}{Theorem}[section]

\newtheorem{Lem}[theorem]{Lemma}
\newtheorem{Claim}[theorem]{Claim}
\newtheorem{Prop}[theorem]{Proposition}

\theoremstyle{definition}
\newtheorem{Def}[theorem]{Definition}

\theoremstyle{remark}

\numberwithin{equation}{section}

\newcommand{\bb}{\mathbb}
\newcommand{\ms}{\mathscr}
\newcommand{\mr}{\mathrm}
\newcommand{\frk}{\mathfrak}

\usepackage{hyperref}
\begin{document}

\title{The $\mr U(1)$-Kepler Problems}
\author{Guowu Meng}

\address{Department of Mathematics, Hong Kong Univ. of Sci. and
Tech., Clear Water Bay, Kowloon, Hong Kong}
% Current address
%\curraddr{Department of Mathematics and Statistics}

\email{mameng@ust.hk}
% \thanks will become a 1st page footnote.

% General info

\date{May 7, 2008}



\begin{abstract} Let $n\ge 2$ be a positive integer.
To each irreducible representation $\sigma$ of $\mr U(1)$, a $\mr
U(1)$-Kepler problem in dimension $(2n-1)$ is constructed and
analyzed. This system is super integrable and when $n=2$ it is
equivalent to a MICZ-Kepler problem. The dynamical symmetry group of
this system is $\widetilde {\mr U}(n, n)$, and the Hilbert space of
bound states ${\ms H}(\sigma)$ is the unitary highest weight
representation of $\widetilde {\mr U}(n, n)$ with highest weight
$$(\underbrace{-1/2, \cdots, -1/2}_n,\; 1/2+\bar \sigma,
\underbrace{1/2, \cdots, 1/2}_{n-1})$$ when $\bar \sigma \ge 0$ or
$$(\underbrace{-1/2, \cdots, -1/2}_{n-1}, -1/2+\bar \sigma,\; \underbrace{1/2, \cdots, 1/2}_n)$$ when $\bar \sigma\le
0$. (Here $\bar\sigma$ is the infinitesimal character of $\sigma$.)
Furthermore, it is shown that the correspondence between $\sigma^*$
(the dual of $\sigma$) and $\ms H(\sigma)$ is the
theta-correspondence for dual pair $(\mr{U}(1), {\mr U}(n,n))$ in
$\mr{Sp}(4n, \bb R)$.
\end{abstract}

\maketitle

\section {Introduction}
The Kepler problem is a well-known physics problem in dimension
three about two bodies which attract each other by a force
proportional to the inverse square of their distance. What is less
known about the Kepler problem is the fact that it is {\em super
integrable}\footnote{A physics model is called {\em super
integrable} if the number of independent symmetry generators is
bigger than the number of degree of freedom. For the Kepler problem,
the degree of freedom is $3$ and the number of independent symmetry
generators is $5$.} at both the classical and the quantum level, and
belongs to a big family of super integrable models. One interesting
such family is the family of MICZ-Kepler problems \cite{MC70,
Z68}.

In our recent research on the Kepler problem \cite{meng05,MZ07,M07,
meng08a}, a dominant theme is the construction of new super
integrable models of Kepler type on the one hand and the exhibition
of the close relationship of these super integrable models with
certain unitary highest weight modules for a real non-compact Lie
group of Hermitian type on the other hand. Depending on the
interests of the readers, one may view this investigation either as
a journey to discover new super integrable models or as an effort to
better understand the geometry of certain unitary highest weight
modules. Here we continue this theme.

In Ref. \cite{meng08a}, a new family of super integrable models of
the Kepler type
--- {\em the $\mr O(1)$-Kepler problems} --- has been constructed and
analyzed, and their intimate relationship with the
theta-correspondence was uncovered. The purpose here is to construct
and analyze their complex analogues
--- {\em the $\mr U(1)$-Kepler problems}.

Recall that, in the construction of the $\mr O(1)$-Kepler problems
in dimension $n$, the canonical bundle
$$
\mr{O}(1)\to \mr{S}^{n-1}\to\bb R\mr{P}^{n-1}
$$
plays a pivotal role. Here the corresponding bundle is
$$
\mr{U}(1)\to \mr{S}^{2n-1}\to\bb C\mr{P}^{n-1}. $$ Later we shall
demonstrate that, a model with $n=2$ constructed in this paper is
equivalent to a MICZ-Kepler problem. That is why the models
constructed here are called the $\mr U(1)$-Kepler problems.

\subsection{Statement of the Main Results}
Before stating our main result, let us fix some notations:
\begin{itemize}
\item $n$ --- an integer which is at least 2;
\item $\sigma$ --- an irreducible representation of $\mr{U}(1)$, it also denotes the underlying representation space of $\sigma$;
\item $\sigma^*$ --- the dual of $\sigma$, it also denotes the underlying representation space of $\sigma^*$;
\item $\bar \sigma$ --- the infinitesimal character of $\sigma$, so it is an integer;
\item $\tilde {\mr U}(n)$ --- the nontrivial double cover of $\mr
U(n)$;
\item ${\mr
U}(n)\tilde\times {\mr U}(n)$ --- the double cover of ${\mr
U}(n)\times {\mr U}(n)$ that corresponds to the homomorphism
$\pi_1({\mr U}(n)\times {\mr U}(n))={\bb Z}\times {\bb Z}\to \bb
Z_2$ sending $(a, b)$ to $a+b\mod 2$;
\item $\tilde{\mr U}(n,n)$ --- the double cover of ${\mr
U}(n,n)$ such that $\left(\tilde{\mr U}(n,n), {\mr U}(n)\tilde\times
{\mr U}(n) \right)$ is the double cover of $\left({\mr U}(n,n), {\mr
U}(n)\times {\mr U}(n) \right)$;
\item $\kappa$ --- a half integer;
\item $l$ --- a nonnegative integer;
\item $\mathcal R_l^\kappa$ --- the highest weight module of
$\tilde{\mr U}(n)$ with highest weight
$$(\kappa+l,\underbrace{\kappa,\cdots,\kappa}_{n-1});$$
\item $\overline{{\mathcal R}_l^\kappa}$ --- the highest weight module of $\tilde{\mr
U}(n)$ with highest weight
$$(\underbrace{-\kappa,\cdots,-\kappa}_{n-1}, -(\kappa+l)).$$
\end{itemize}

We are now ready to state the main results on the $\mr U(1)$-Kepler
problems.
\begin{Thm}\label{T:main} Let $n\ge 2$ be an integer, $\sigma$ an irreducible representation of $\mr{U}(1)$, $\sigma^*$ the dual of $\sigma$,
and $\bar \sigma$ the infinitesimal character of $\sigma$. For
integer $l\ge 0$, the highest weight modules of $\tilde{\mr U}(n)$
with highest weight $(1/2+l,\underbrace{1/2,\cdots,1/2}_{n-1})$ (
$(\underbrace{-1/2,\cdots,-1/2}_{n-1}, -(1/2+l))$ respectively) is
denoted by ${\mathcal R}_l^{1/2}$ ($\overline{{\mathcal R}_l^{1/2}}$
respectively).

For the $(2n-1)$-dimensional ${\mr U}(1)$-Kepler problem with
magnetic charge $\sigma$, the following statements are true:

1) The bound state energy spectrum is
$$
E_I=-{1/2\over (I+{n+|\bar\sigma|\over 2})^2}
$$ where $I=0$, $1$, $2$, \ldots

2) There is a natural unitary action of $\widetilde {\mr U}(n, n)$
on the Hilbert space ${\ms H}(\sigma)$ of bound states, which
extends the manifest unitary action of $\mr{U}(n)$ --- the diagonal
of ${\mr U}(n)\tilde\times {\mr U}(n)$. In fact, ${\ms H}(\sigma)$
is the unitary highest weight module of $\widetilde {\mr U}(n, n)$
with highest weight
$$(\underbrace{-1/2, \cdots, -1/2}_n,\; 1/2+\bar \sigma,
\underbrace{1/2, \cdots, 1/2}_{n-1})$$ when $\bar \sigma \ge 0$ or
$$(\underbrace{-1/2, \cdots, -1/2}_{n-1}, -1/2+\bar \sigma,\; \underbrace{1/2, \cdots, 1/2}_n)$$ when $\bar \sigma\le 0$.

3) When restricted to the maximal compact subgroup ${\mr
U}(n)\tilde\times {\mr U}(n)$, the above action yields the following
orthogonal decomposition of $\ms H(\sigma)$:
$$
{\ms H}(\sigma)=\hat\bigoplus _{I=0}^\infty\,{\ms H}_I(\sigma)
$$ where, as an irreducible ${\mr U}(n)\tilde\times
{\mr U}(n)$-module,
$${\ms H}_I(\sigma)\cong \left\{
\begin{array}{cc}
\overline{{\mathcal R}_I^{1/2}}\otimes {\mathcal R}_{I+|\bar
\sigma|}^{1/2} &
\mbox{ when $\bar\sigma\ge 0$}\\
\\
\overline {{\mathcal R}_{I+|\bar \sigma|}^{1/2}}\otimes {\mathcal
R}_I^{1/2} & \mbox{ when $\bar\sigma\le 0$}.
\end{array}\right.
$$

4) ${\ms H}_I(\sigma)$ in part 3) is the energy eigenspace with
eigenvalue $E_I$ in part 1).

5) The correspondence between $\sigma^*$ and $\ms H(\sigma)$ is the
theta-correspondence\footnote{See Ref. \cite{Howe} for details on
reductive dual pairs and theta-correspondence.} for dual pair $(\mr
U(1), \mr {U}(n, n))\subset \mr{Sp}(4n, \bb R)$.
\end{Thm}
For readers who are familiar with the Enright-Howe-Wallach
classification diagram \cite{EHW82} for the unitary highest weight
modules, we would like to point out that, the unitary highest weight
module identified in part 2) of this theorem occurs at the rightmost
nontrivial reduction point of the classification diagram. Note that,
part 3) of the above theorem is a \emph{multiplicity free} $K$-type
formula.

In section \ref{S:model}, we introduce the models and show that when
$n=2$ they are equivalent to the MICZ-Kepler problems. In section
\ref{S:analysis}, we give a detailed analysis of the models and
finish the proof of Theorem \ref{T:main}.

\section{The models}\label{S:model}
Let $n\ge 2$ be an integer, $\bb C^n_*=\bb C^n\setminus\{0\}$, and
$\sigma$ an irreducible unitary representation of $\mr U(1)$. Denote
by $\bar \sigma$ be the infinitesimal character of $\sigma$.

Consider the principal bundle
$$
{\mr U}(1)\to \bb C^n_*\to \widetilde{\bb CP^n}
$$
where $\widetilde{\bb CP^n}\cong \bb R_*\times {\bb C}P^{n-1}$ is
the quotient space of $\bb C^n_*$ under the equivalence relation
$Z\sim \alpha\cdot Z$, $\alpha\in \mr{U}(1)$. We shall assume the
Euclidean metric on $\bb C^n_*$ and the resulting quotient
Riemannian metric on $\widetilde{\bb CP^n}$.

We use $(\rho, \Phi)$ to denote the polar coordinates on
$\widetilde{\bb CP^n}$ and $\gamma_\sigma$ to denote the vector
bundle associated to representation $\sigma$. Then $\rho([Z])=|Z|$,
$\Phi$ is a coordinate on $\bb CP^{n-1}$, and $\gamma_\sigma$ is a
hermitian line bundle over $\widetilde{\bb CP^n}$ with a natural
hermitian connection $\mathcal A$. Here is a technical lemma which
will be used later.
\begin{Lem}\label{1stLem}
1) Let $Z\in \bb C^n_*$ and $\rho=|Z|=\sqrt{\bar Z\cdot Z}$. Then we
have the following identity for the Euclidean metric on $\bb C^n_*$:
$$
|dZ|^2=d\rho^2+\rho^2\left(ds^2_{FS}+\left({Im(\bar Z\cdot dZ)\over
|Z|^2}\right)^2\right)
$$ where ``$Im$'' in $Im(\bar Z\cdot dZ)$ stands for ``the imaginary part
of", and $ds^2_{FS}$ is the Fubini-Study metric\footnote{It is the
quotient metric on $\mr{S}^{2n-1}/\mr{U}(1)$. On ${\bb CP}^1={\mr
S}^2$, it is $1\over 4$ times the standard round metric.} on ${\bb
C}P^{n-1}$, i.e.,
\begin{eqnarray}
ds^2_{FS}={|dZ|^2\over |Z|^2}-{|Z\cdot d\bar Z|^2\over |Z|^4}.
\end{eqnarray} Consequently, the quotient metric on $\widetilde{\bb
CP^n}$ is
\begin{eqnarray}
ds^2_{\widetilde{\bb CP^n}}=d\rho^2+\rho^2 ds^2_{FS}.
\end{eqnarray}

2) If we take the invariant Riemannian metric on $\mr{U}(n)$ as the
restriction of the Euclidean metric on $\bb C^{n^2}$, then the
resulting quotient metric on $\bb CP^{n-1}={\mr{U}(n)\over
\mr{U}(n-1)\times\mr{U}(1)}$ is twice of the Fubini-Study metric.
Consequently,
\begin{eqnarray}\label{LapFor}
\Delta_{\mathcal A}|_{\bb CP^{n-1}}=2\left(c_2[{\mr
U}(n)]-c_2[{\mr U}(1)]|_\sigma\right)
\end{eqnarray} where
$\Delta_{\mathcal A}|_{\bb CP^{n-1}}$ is the (non-negative) Laplace
operator acting on sections of $\gamma_\sigma|_{\bb CP^{n-1}}$,
$c_2[{\mr U}(n)]$ is the Casimir operator of ${\mr U}(n)$, and
$c_2[{\mr U}(1)]|_\sigma$ is the value of the Casimir operator of
${\mr U}(1)$ at $\sigma$.

\end{Lem}
\begin{proof} 1) Let $Z=(Z_1, \cdots. Z_n)\in \bb C^n_*$. The Euclidean metric on
$\bb C^n_*$ is just $|dZ|^2=dZ\cdot d\bar Z$. Since
$\rho^2=|Z|^2=Z\cdot \bar Z$, $2\rho d\rho=Z\cdot d\bar Z+\bar
Z\cdot dZ$, and $$ d\rho^2={2|Z\cdot d\bar Z|^2+(Z\cdot d\bar
Z)^2+(\bar Z\cdot dZ)^2\over 4|Z|^2}.
$$ A computation shows that
$$
|dZ|^2=d\rho^2+\rho^2\left(ds^2_{FS}+\left({Im(\bar Z\cdot dZ)\over
|Z|^2}\right)^2\right).
$$
The right $\mr{U}(1)$ action on $\bb C^n_*$ generates a vector field
on $\bb C^n_*$ whose value at point $Z_0\in \bb C^n_*$ is $(Z_0,
iZ_0)\in T_{Z_0} \bb C^n_*$. Let $V_{Z_0}$ be the orthogonal
complement of $\mr{span}_{\bb R}\{iZ_0\}$ in ${\bb C}^n$($=\bb
R^{2n}$), then
$$
V_{Z_0}=\left\{W\in \bb C^{n}\mid Im(\bar Z_0\cdot W)=0 \right\}.
$$

To finish the proof of part 1), we just need to check that quadratic
form
$$\left.\left(Im(\bar Z\cdot dZ)\right)^2\right|_{Z_0}$$ vanishes
when restricted to $\{Z_0\}\times V_{Z_0}$. Indeed, if $v:=(Z_0,
W)\in \{Z_0\}\times V_{Z_0}$, then
$$
\left.\left(Im(\bar Z\cdot dZ)\right)^2\right|_{Z_0}(v, v)=(Im(\bar
Z_0\cdot W))^2=0.
$$

2) The Euclidean metric on $\bb C^{n^2}$ is $\mr{tr}(dM^\dag\, dM)$,
so the Riemannian metric on $\mr{U}(n)$ is
$ds^2_{\mr{U}(n)}=\mr{tr}(dg^\dag\, dg)$. Since the action of $\mr
U(n)$ on $\bb CP^{n-1}={\mr{U}(n)\over \mr{U}(n-1)\times\mr{U}(1)}$
is transitive and both the Fubini-Study metric and the quotient
metric here are $\mr{U}(n)$ invariant, we just need to verify the
statement at point $p:=[1, 0,\cdots ,0]\in \bb CP^{n-1}$.

In the following, if $X$ is a Riemannian manifold, we shall use
$\langle, \rangle|_X$ to denote the Riemannian inner product on a
tangent space of $X$.

Let $I\in \mr U(1)$ be the identity matrix. Let $u, v\in
T_I\mr{U}(n)$ be two tangent vectors which are orthogonal to
$T_I(\mr{U}(n-1)\times\mr{U}(1))$. Then, there are $a, b\in \bb
C^{n-1}$ such that
$$
u=\left(I, \left( \begin{matrix}0 & -a^\dag\cr a & 0\end{matrix}
\right)\right), \quad v=\left(I, \left( \begin{matrix}0 & -b^\dag
\cr b & 0\end{matrix} \right)\right),
$$ so $\langle u, v\rangle|_{\mr{U}(n)}=a^\dag b+b^\dag a=2Re(a^\dag b)$.
Let $$\tilde p=(1, \underbrace{0, \cdots, 0}_{n-1})^T\in
\mr{S}^{2n-1},$$ and $\tilde u$, $\tilde v$ be the image of $u$, $v$
respectively under the linearlization at $I$ of the quotient map
\begin{eqnarray}
{\mr{U}}(n) &\to & \mr {S}^{2n-1}\cr A&\mapsto & A\tilde
p\;.\nonumber
\end{eqnarray} Then
$$
\tilde u=\left(\tilde p, \left( \begin{matrix}0 \cr a \end{matrix}
\right)\right), \quad \tilde v=\left(\tilde p, \left(
\begin{matrix}0 \cr b \end{matrix} \right)\right),
$$ so $\langle \tilde u, \tilde v\rangle|_{\mr{S}^{2n-1}}=Re(a^\dag b)$.
Therefore,
\begin{eqnarray}\label{2wice}
\langle u, v\rangle|_{\mr{U}(n)}=2\langle \tilde u, \tilde
v\rangle|_{\mr{S}^{2n-1}}.
\end{eqnarray}

Let $\bar u$, $\bar v$ be the image of $u$, $v$ respectively under
the linearlization at $I$ of the quotient map
\begin{eqnarray} {\mr{U}}(n)
&\to& \bb CP^{n-1}\cr A&\mapsto&[A\tilde p]\;.\nonumber
\end{eqnarray}
One can see that $u$, $v$ are respectively the horizontal lift of
$\bar u$, $\bar v$, then, by definition,
$$ \fbox{$\langle
\bar u, \bar v\rangle_q=\langle u, v\rangle|_{\mr{U}(n)}$\;.}
$$
Here $ \langle \;, \rangle_q$ denote the quotient metric on
${\bb CP}^{n-1}={\mr{U}(n)\over \mr{U}(n-1)\times \mr{U}(1)}$.

On the other hand, one can see that $\tilde u$ and $\tilde v$ are
respectively the horizontal lift of $\bar u$ and $\bar v$ in fiber
bundle $\mr{S }^{2n-1} \to \bb CP^{n-1}$, then, by definition,
$$\fbox{$ \langle \bar u, \bar v\rangle_{FS}=\langle \tilde u, \tilde
v\rangle|_{\mr{S}^{2n-1}}\; .$}
$$

Eq. (\ref{2wice}) then implies that $ \langle \bar u, \bar
v\rangle_q= 2\langle \bar u, \bar v\rangle_{FS}$. In view of the
fact that the Laplace operator gets multiplied by $1/a$ if the
Riemannian metric gets multiplied by a number $a$, Eq.
(\ref{LapFor}) follows from Lemma \ref{LemAp} in appendix
\ref{appendix}.

\end{proof}
We are now ready to introduce the notion of  $\mr U(1)$-Kepler
problems.

\begin{Def}\label{Def:1st} Let $n\ge 2$ be an integer and
$\sigma$ an irreducible representation of $\mr U(1)$. The $\mr
U(1)$-Kepler problem in dimension $(2n-1)$ with magnetic charge
$\sigma$ is the quantum mechanical system for which the wave
functions are smooth sections of $\gamma_\sigma$, and the
hamiltonian is
\begin{eqnarray}\label{H:1st}
H=-{1\over 8\rho}\Delta_{\mathcal A}{1\over \rho}+{\bar
\sigma^2+(n-{5\over 4})\over 8\rho^4}-{1\over \rho^2}
\end{eqnarray}
where $\Delta_{\mathcal A}$ is the (non-positive) Laplace operator
on $\widetilde{\bb CP^n}$ twisted by $\gamma_\sigma$, $\bar\sigma$
is the infinitesimal character of $\sigma$, and $\rho([Z])=|Z|$.
\end{Def}

Note that $\bb R^3_*$ and $\widetilde{\bb CP^2}$ are diffeomorphic.
We use $(r, \Theta)$ to denote the polar coordinates on $\bb R^3_*$
and $(\rho, \Phi)$ to denote the polar coordinates on
$\widetilde{\bb CP^2}$. Let $\pi$: $\widetilde{\bb CP^2}\to \bb
R^3_*$ be the diffeomorphism such that $\pi(\rho, \Phi)=(\rho^2,
\Phi)$, then
$$
\pi^*(dr^2+r^2\,d\Theta^2)=4\rho^2 (d\rho^2+\rho^2\,ds^2_{FS})\quad
\mbox{and}\quad \pi^*(\mr{vol}_{{\bb
R^3}_*})=(2\rho)^3\mr{vol}_{\widetilde{\bb CP^2}}.
$$

Let $\gamma(\bar \sigma)$ be the pullback of $\gamma_\sigma$ by
$\pi^{-1}$. It is clear that $\gamma(\bar \sigma)$ is a hermitian
line bundle over $\bb R^3_*$ with a natural hermitian connection
$A$. Recall from Refs. \cite{MC70,Z68, meng05} that \emph{the MICZ
Kepler problem with magnetic charge $\mu\in {1\over 2}\bb Z$} is the
quantum mechanical system for which the wave functions are smooth
sections of $\gamma(\bar \sigma)$ ($\mu=\bar \sigma/2$), and the
hamiltonian is
$$\hat h_\mu=-{1\over 2}\Delta_A+{\mu^2\over 2r^2}-{1\over r}$$ where $\Delta_A$ is
the (non-positive) Laplace operator on $\bb R^3_*$ twisted by
$\gamma(\bar \sigma)$ and $r(x)=|x|$. We are now ready to state the
following
\begin{Prop}\label{equiv} The $\mr U(1)$-Kepler problem in dimension three with magnetic charge
$\sigma$ is equivalent to the MICZ-Kepler problem with magnetic
charge ${\bar \sigma\over 2}$.
\end{Prop}
\begin{proof}
Let $\Psi_i$ ($i=1$ or $2$) be a wave-section for the MICZ-Kepler
problem with magnetic charge $\bar \sigma/2$, and
$$\psi_i(\rho, \Phi):=(2\rho)^{3\over 2}\,\pi^*(\Psi_i)(\rho,
 \Phi)=(2\rho)^{3\over 2}\,\Psi_i(\rho^2, \Phi).$$ Then it is not
hard to see that
$$\displaystyle \int_{\widetilde{\bb
CP^2}}\overline{\psi_1}\psi_2\,\mr{vol}_{\widetilde{\bb
CP^2}}=\displaystyle \int_{\widetilde{\bb
CP^2}}\overline{\pi^*(\Psi_1)}\pi^*(\Psi_2)\, \pi^*(\mr{vol}_{\bb
R^3_*})=\displaystyle \int_{\bb R^3_*}\overline{\Psi_1}
\Psi_2\,\mr{vol}_{\bb R^3_*}$$ and
\begin{eqnarray}\displaystyle
\displaystyle\int_{\widetilde{\bb CP^2}}\overline{\psi_1}H\psi_2\,
\mr{vol}_{\widetilde{\bb CP^2}} &=&
\displaystyle\int_{\widetilde{\bb
CP^2}}\overline{\pi^*(\Psi_1)}\,{1\over \rho^{3\over
2}}H\rho^{3\over 2}\,\pi^*(\Psi_2) \,\pi^*(\mr{vol}_{\bb
R^3_*})\cr&=& \displaystyle\int_{\bb R^3_*}\overline{\Psi_1}\,\hat
h_{\bar\sigma/2}\,\Psi_2 \,\mr{vol}_{\bb R^3_*}. \nonumber
\end{eqnarray}Here we have used the fact that
\begin{eqnarray}
{1\over \rho^{3\over 2}}H\rho^{3\over 2} &=& -{1\over
8\rho^{5/2}}\left({1\over
\rho^2}\partial_\rho\rho^2\partial_\rho-{1\over \rho^2}\Delta_
{\mathcal A}|_{\bb CP^1}\right)\rho^{1/2}+{\bar\sigma^2+3/4\over
8\rho^4}-{1\over \rho^2}\cr &=& -{1\over 2}\left({1\over
r^2}\partial_rr^2\partial_r-{1\over r^2}\Delta_A|_{\mr
S^2}\right)+{\bar \sigma^2\over 8 r^2}-{1\over r}\cr &=&-{1\over
2}\Delta_A+{(\bar \sigma/2)^2\over 2r^2}-{1\over r}=\hat
h_{\bar\sigma/2}.\nonumber
\end{eqnarray}

\end{proof}

\section{The dynamical symmetry analysis}\label{S:analysis}
Let $\psi$ be an eigensection of $H$ in Eq. (\ref{H:1st}) with
eigenvalue $E$, so $\psi$ is square integrable with respect to
volume form $\mr{vol}_{\widetilde{\bb CP^n}}$, and
\begin{eqnarray}\label{E:eigenvalue}
\left(-{1\over 8\rho}\Delta_{\mathcal A}{1\over \rho}+{\bar
\sigma^2+(n-{5\over 4})\over 8\rho^4}-{1\over
\rho^2}\right)\psi=E\psi.
\end{eqnarray}
We shall solve this eigenvalue problem by separating the angles from
the radius.

In view of the branching rule for $({\mr U}(n), {\mr U}(n-1)\times
{\mr U}(1))$, the Frobenius reciprocity theorem implies that, as
modules of $\mr {U}(n)$,
\begin{eqnarray}\label{Decomposition}
L^2(\gamma_\sigma|_{{\bb C}P^{n-1}})=\hat\bigoplus_{p,q\in \bb N}
{\ms R}_{p, q}^\sigma \end{eqnarray} where ${\ms R}_{p,q}^\sigma$ is
the irreducible and unitary representation of $\mr{U}(n)$ with the
highest weight $(p, 0, \cdots, 0, -q)$ subject to condition
$p-q=\bar \sigma$. For nonnegative integers $p$, $q$ with $p-q=\bar
\sigma$, we introduce the notation
$$
l=\left\{
\begin{array}{ll}
q & \mbox{if $\bar\sigma \ge 0$}\\
\\
p & \mbox{if $\bar\sigma \le 0$},
\end{array}\right.
$$
and rewrite ${\ms R}_{p,q}^\sigma$ as ${\ms R}_l(\sigma)$, then Eq.
(\ref{Decomposition}) becomes
\begin{eqnarray}
L^2(\gamma_\sigma|_{{\bb C}P^{n-1}})=\hat\bigoplus_{l=0}^\infty {\ms
R}_l(\sigma).
\end{eqnarray}

Let $\{Y_{{l\bf m}}\mid {\bf m}\in {\mathcal I}(l)\}$ be a minimal
spanning set for ${\ms R}_l(\sigma)$. Write $\psi(x)=\tilde
R_{kl}(\rho)Y_{l\bf m}(\Phi)$. After separating out the angular
variables with the help of Eq. (\ref{LapFor}) in Lemma \ref{1stLem},
Eq. (\ref{E:eigenvalue}) becomes
\begin{eqnarray}
\left(-{1\over
8\rho^{2n-1}}\partial_\rho\rho^{2n-2}\partial_\rho{1\over
\rho}+{(l+{|\bar\sigma|\over 2})^2+(n-1)(l+{|\bar\sigma|\over
2})+{1\over 4}(n-{5\over 4})\over 2\rho^4}-{1\over
\rho^2}\right)\tilde R_{kl}=E\tilde R_{kl}.\nonumber
\end{eqnarray} where $\tilde R_{kl}\in L^2({\bb R}_+, \rho^{2n-2}\,d\rho)$.
Let $R_{k(l+{|\bar\sigma|\over 2})}(t)=\tilde R_{kl}(\sqrt
t)/t^{3/4}$, then we have $ R_{k(l+{|\bar\sigma|\over 2})}\in
L^2({\bb R}_+, t^n\,dt)$ and
\begin{eqnarray}
\left( -{1\over 2t^n}\partial_t t^n\partial_t+{(l+{|\bar\sigma|\over
2})^2+(n-1)(l+{|\bar\sigma|\over 2})\over 2t^2}-{1\over t}\right)
R_{k(l+{|\bar\sigma|\over 2})}=E R_{k(l+{|\bar\sigma|\over
2})}.\nonumber
\end{eqnarray}
By quoting results from appendix A in Ref. \cite{meng08a}, we have
\begin{eqnarray}
E_{k(l+{|\bar\sigma|\over 2})}=-{1/2\over (k+l+{|\bar\sigma|+n\over
2}-1)^2}
\end{eqnarray} where $k=1, 2, 3, \cdots$.
Let $I=k-1+l$, then the {\em bound energy spectrum} is
\begin{eqnarray}
E_I=-{1/2\over (I+{n+|\bar\sigma|\over 2})^2} \end{eqnarray} where
$I=0, 1, 2, \cdots$. Part 1) of Theorem \ref{T:main} is proved.
Moreover, since $\tilde R_{kl}(\rho)=\rho^{3\over
2}R_{k(l+{|\bar\sigma|\over 2})}(\rho^2)$, we have
$$
\tilde R_{kl}(\rho)=c(k,l+{|\bar\sigma|\over
2})\rho^{2l+|\bar\sigma|+3/2}L^{2l+|\bar\sigma|+n-1}_{k-1}\left({2\rho^2\over
I+{n+|\bar\sigma|\over 2}}\right)\exp\left(-{\rho^2\over
I+{n+|\bar\sigma|\over 2}}\right).
$$

For each integer $I\ge 0$, we let ${\ms H}_I(\sigma)$ be the linear
span of
$$
\{\tilde R_{kl}Y_{l{\bf m}}\mid {\bf m}\in {\mathcal I}(l),
k-1+l=I\},
$$ then
\begin{eqnarray}\label{energyE}
{\ms H}_I(\sigma)\cong \bigoplus_{l=0}^{I} {\ms R}_l(\sigma)
\end{eqnarray} is the eigenspace of $H$ with eigenvalue $E_I$, and the
Hilbert space of bound states admits the following orthogonal
decomposition into the eigenspaces of $H$:
$$
{\ms H}(\sigma)=\hat\bigoplus_{I=0}^\infty {\ms H}_I(\sigma).
$$
Part 4) of Theorem \ref{T:main} is then clear. We shall show that,
$\mr U(n)\tilde\times \mr U(n)$ acts on ${\ms H}_I(\sigma)$, and
$${\ms H}_I(\sigma)\cong \left\{
\begin{array}{cc}
\overline{{\mathcal R}_I^{1/2}}\otimes {\mathcal R}_{I+|\bar
\sigma|}^{1/2} &
\mbox{ when $\bar\sigma\ge 0$}\\
\\
\overline {{\mathcal R}_{I+|\bar \sigma|}^{1/2}}\otimes {\mathcal
R}_I^{1/2} & \mbox{ when $\bar\sigma\le 0$}
\end{array}\right.
$$ as irreducible representations of $\mr
U(n)\tilde\times \mr U(n)$. To do that, we need to twist the Hilbert
space of bound states and the energy eigenspaces.
\subsection{Twisting}
Let $n_I=I+{n+|\bar\sigma|\over 2}$ for each integer $I\ge 0$. For
each $\psi_I\in {\ms H}_I$, as in Refs. \cite{Barut71,MZ07}, we
define its twist $\tilde \psi_I$ by the following formula:
\begin{eqnarray}
\tilde \psi_I(Z)=c_I{1\over |Z|^{3\over 2}}\psi_I(\sqrt{n_I\over
2}[Z])
\end{eqnarray} where $c_I$ is the unique constant such that
$$\int_{{\bb
C^n_*}}|\tilde \psi_I|^2=\int_{\widetilde{\bb CP^n}}|\psi_I|^2.$$

We use $\tilde {\ms H}_I(\sigma)$ to denote the span of all such
$\tilde\psi_I$'s, $\tilde {\ms H}(\sigma)$ to denote the Hilbert
space direct sum of $\tilde{\ms H_I}(\sigma)$. We write the linear
map sending $\psi_I$ to $\tilde\psi_I$ as
\begin{eqnarray}\label{taumap}
\tau:\quad {\ms H}(\sigma)\to \tilde{\ms H}(\sigma).
\end{eqnarray}
Then $\tau$ is a linear isometry.

Since
$$
H\psi_I=E_I\psi_I,
$$
after re-scaling: $[Z]\to \sqrt{n_I\over 2}[Z]$, we have
$$
\left(-{(2/n_I)^2\over 8r}\Delta_{\mathcal A}{1\over
r}+(2/n_I)^2{\bar \sigma^2+(n-{5\over 4})\over 8r^4}-{2/n_I\over
r^2}\right)\psi_I(\sqrt{n_I\over 2}[Z])=E_I\psi_I(\sqrt{n_I\over
2}[Z]),
$$ where $r=|Z|$. Multiplying by $(n_Ir)^2$, we obtain
\begin{eqnarray}
{1\over r^{3/2}}\left(-{r\over 2}\left(\Delta_{\mathcal A}-{\bar
\sigma^2+(n-{5\over 4})\over r^2}\right){1\over
r}-{2n_I}\right)r^{3/2}\tilde\psi_I(Z)&= &
n_I^2E_Ir^2\tilde\psi_I(Z)\cr &=&-{1\over 2}
r^2\tilde\psi_I(Z).\nonumber
\end{eqnarray} Applying Lemma \ref{LemAp}, the previous equation becomes
\begin{eqnarray}
-{1\over 2} r^2\tilde\psi_I&=&\left(-{1\over 2}\left({1\over
r^{2n-{3\over 2}}}\partial_r r^{2n-2}\partial_r r^{1\over
2}-{\Delta\mid _{{\mr S}^{2n-1}}+n-{5\over 4}\over
r^2}\right)-{2n_I}\right)\tilde\psi_I\cr &=& \left(-{1\over
2}\left({1\over r^{2n-1}}\partial_r r^{2n-1}\partial_r-{\Delta\mid
_{{\mr S}^{2n-1}}\over r^2}\right)-{2n_I}\right)\tilde\psi_I\cr &=&
\left(-{1\over 2}\Delta-2n_I\right)\tilde \psi_I, \nonumber
\end{eqnarray} where $\Delta$ is the (negative definite) standard Laplace operator on $\bb
C^n$, and $\tilde \psi_I$ is viewed as a $\sigma$-valued function on
$\bb C^n_*$. Then
\begin{eqnarray}\label{Osc}\left(-{1\over 2}\Delta+{1\over 2} r^2\right)\tilde
\psi_I=(2I+|\bar\sigma|+n)\tilde \psi_I.
\end{eqnarray}

Denote the collection of all smooth sections of $\gamma_\sigma$ by
$C^\infty(\gamma_\sigma)$, and the collection of all smooth
complex-valued functions on $\bb C^n$ by $C^\infty(\bb C^n_*)$. Note
that an element in $\sigma^*\otimes C^\infty(\gamma_\sigma)$ can be
viewed as a smooth map $F$: $\bb C^n_*\to \mr{End} (\sigma)$ such
that $F(x\cdot g^{-1})=\rho_\sigma(g)\circ F(x)$ for any $g\in
\mr{U}(1)$. Denote by $\mr{Tr}\,F$ the map sending $x\in \bb C^n$ to
the trace of $F(x)$ and by $\mr{Tr}$ the linear map sending
$F\in\sigma^*\otimes C^\infty(\gamma_\sigma)$ to $\mr{Tr}\,F\in
C^\infty(\bb C^n_*)$.

The left $\mr{U}(n)$-action on $\bb C^n$ induces natural actions on
both $C^\infty(\gamma_\sigma)$ (hence on $\sigma^*\otimes
C^\infty(\gamma_\sigma)$) and $C^\infty(\bb C^n_*)$ with respect to
which $\mr{Tr}$ is equivariant: For any $g\in\mr{U}(n)$,
$$
\mr{Tr}\left(L_g^*F\right)(x)=\mr{Tr}\left(L_g^*F(x)\right)=\mr{Tr}\left(F(g\cdot
x )\right)=\mr{Tr}\,F(g\cdot x)=L_g^*\left(\mr{Tr}\,F\right)(x).
$$

The right $\mr{U}(1)$-action on $\bb C^n$ induces a natural action
on $C^\infty(\bb C^n_*)$. The action of ${\mr U}(1)$ on $\sigma^*$
yields a natural $\mr{U}(1)$-action on $\sigma^*\otimes
C^\infty(\gamma_\sigma)$ for which
$$
(g\cdot F) (x)=F(x)\circ \rho_\sigma(g^{-1})\quad\mbox{for any
$g\in\mr{U}(1)$}.
$$
It is clear that $\mr{Tr}$ is also $\mr{U}(1)$-equivariant: For any
$g\in\mr{U}(1)$,
\begin{eqnarray}
\mr{Tr}\left(g\cdot F\right)(x)&=& \mr{Tr}\left((g\cdot F)(x)\right)
\cr &= & \mr{Tr}\left(F(x)\circ\rho_\sigma(g^{-1}) \right)\cr &=
&\mr{Tr}\left(\rho_\sigma(g^{-1})\circ F(x) \right)\cr &=
&\mr{Tr}\left(F(x \cdot g)\right)=\mr{Tr}\,F(x \cdot g)\cr
&=&R_g^*(\mr{Tr}\,F)(x)\nonumber
\end{eqnarray} where $R_g$ stands for the right action of $g$
on $\bb C^n_*$.

On $\sigma^*\otimes C^\infty(\gamma_\sigma)$, there is a natural
inner product:
$$
\langle\alpha\otimes \psi, \beta\otimes \phi\rangle=\langle\alpha,
\beta \rangle\int_{\bb C^n_*} \langle \psi, \phi\rangle.
$$
On $C^\infty(\bb C^n_*)$, there is a natural inner product:
$$
\langle f, g\rangle=\int_{\bb C^n_*} \bar f g.
$$
It is easy to see that $\mr{Tr}$ is an isometry: if we let $v$ be an
orthonormal basis of $\sigma$, $\hat v$ be the dual of $v$, then an
element in $\sigma^*\otimes C^\infty(\gamma_\sigma)$ can be written
as $f\,v\hat v$. So
\begin{eqnarray}
\langle \mr{Tr}\, (f\,v\hat v), \mr{Tr}\, (g\,v\hat v)\rangle &=
&\langle f, g\rangle=\int_{\bb C^n} \bar f g\cr &=& \langle\hat v,
\hat v\rangle\int_{\bb C^n} \langle fv, gv\rangle\cr &=& \langle
f\,v\hat v, g\,v\hat v\rangle.\nonumber
\end{eqnarray}

Since $\mr{Tr}$ is an isometry, it must be injective. Actually, we
have
\begin{Prop} $\mr{Tr}$ induces a Hilbert space isomorphism
\begin{eqnarray}\label{corres}
\mr{Tr}_*:\quad \hat \bigoplus_{\sigma^* \in \widehat {\mr
{U}(1)}}\sigma^*\otimes\tilde{\ms H}(\sigma) &\cong&
L^2\left(\bb C^n_*\right)=L^2\left(\bb C^n\right) \\
\oplus_\sigma F_\sigma & \mapsto &\oplus_\sigma \mr{Tr}\,
F_\sigma\nonumber
\end{eqnarray} which is equivariant with respect to both the
$\mr{U}(n)$-actions and the $\mr{U}(1)$-actions.
\end{Prop}
\begin{proof}

Let ${\mathcal H}_k $ be the $k$-th energy eigenspace of the
$2n$-dimensional isotropic harmonic oscillator with hamiltonian
$-{1\over 2}\Delta+{1\over 2} r^2$. In view of Eq. (\ref{Osc}), we
have the following map
\begin{eqnarray}\label{claimEq}
\iota:=\oplus\mr{Tr}:\quad
\bigoplus_{2I+|\bar\sigma|=k}\sigma^*\otimes\tilde{\ms
H}_I(\sigma)\longrightarrow {\mathcal H}_k.\end{eqnarray}
\begin{Claim} $\iota$ is an isomorphism.
\end{Claim}
\begin{proof}Since $\iota$ is injective, one just needs to verify the {\em
dimension equality}:
$$
\sum_{2I+|\bar\sigma|=k} \dim\sigma^* \cdot \dim \tilde{\ms
H}_I(\sigma)= \dim {\mathcal H}_k.
$$ In view of Eq. (\ref{energyE}),
\begin{eqnarray}
\dim \tilde{\ms H}_I(\sigma)&= &\dim {\ms
H}_I(\sigma)=\sum_{l=0}^{I} \dim {\ms R}_l(\sigma)\cr
&=&\sum_{{p+q-|p-q|\over 2}\le I} \dim {\ms
R}_{p,q}^\sigma.\nonumber
\end{eqnarray} Using the dimension formula in representation theory to compute $\dim {\ms
R}_{p,q}^\sigma$, one arrives at the following explicit form of the
dimension equality:
\begin{eqnarray}\fbox{$ {\sum_{{p+q-|p-q|\over 2}\le
I}^{2I+|p-q|=k}}{p+q+n-1\over n-1}\left(\begin{matrix}p+n-2\cr
n-2\end{matrix}\right)\left(\begin{matrix}q+n-2\cr
n-2\end{matrix}\right)=\left(\begin{matrix}2n+k-1\cr
2n-1\end{matrix}\right)$.}\nonumber
\end{eqnarray}
To prove this identity, we use the method of generating functions.
The details are as follows:
\begin{eqnarray}
& & \sum_{k=0}^\infty {\sum_{{p+q-|p-q|\over 2}\le
I}^{2I+|p-q|=k}{p+q+n-1\over n-1}}\left(\begin{matrix}p+n-2\cr
n-2\end{matrix}\right)\left(\begin{matrix}q+n-2\cr
n-2\end{matrix}\right)t^k \cr & = & \sum_{p,q\ge 0}{t^{p+q}\over
1-t^2}{p+q+n-1\over n-1}\left(\begin{matrix}p+n-2\cr
n-2\end{matrix}\right)\left(\begin{matrix}q+n-2\cr
n-2\end{matrix}\right) \cr & = & \sum_{p,q\ge 0}{t^{p+q}\over
1-t^2}\left[\left(\begin{matrix}p+n-1\cr
n-1\end{matrix}\right)\left(\begin{matrix}q+n-2\cr
n-2\end{matrix}\right)+\left(\begin{matrix}p+n-2\cr
n-2\end{matrix}\right)\left(\begin{matrix}q+n-2\cr
n-1\end{matrix}\right)\right]\cr &=&{1\over
1-t^2}\left((1-t)^{-2n+1}+t(1-t)^{-2n+1}\right)= (1-t)^{-2n}\cr &=&
\sum_{k\ge 0}\left(\begin{matrix}2n+k-1\cr
2n-1\end{matrix}\right)t^k.\nonumber
\end{eqnarray}
\end{proof}
As a consequence of the above claim, we have the following Hilbert
space isomorphism:
\begin{eqnarray}
\hat \bigoplus_{\sigma^* \in \widehat {\mr
{U}(1)}}\sigma^*\otimes\tilde{\ms H}(\sigma)&= &\hat
\bigoplus_{\sigma \in \widehat {\mr U(1)}, I\ge 0}
\sigma^*\otimes\tilde{\ms H}_I(\sigma)\cr
&\buildrel\mr{Tr}_*\over\cong&\hat \bigoplus_{k=0}^\infty {\mathcal
H}_k\cr &=&L^2\left(\bb C^n_*\right).\nonumber
\end{eqnarray}

\end{proof}

\subsection{Proof of Theorem \ref{T:main} }
Parts 1) and 4) have been proved.  Let $(q^i, p_i)$'s be the
canonical symplectic coordinates on $T^*\bb C^n$, in terms of which,
the canonical symplectic form can be written as
$$\omega=dq^i\wedge dp_i\,.$$
The complex structure $I$ on $\bb C^n$ is a diffeomorphism from $\bb
C^n$ to itself, so $T^*I$ is a symplectomorphism from $(T^*\bb C^n,
\omega)$ to itself. Introducing $z^i=q^{2i-1}+\sqrt{-1}q^{2i}$ and
$w_i=p_{2i-1}+\sqrt{-1}p_{2i}$, one can check that $T^*I(z^i,
w_i)=\sqrt{-1}(z^i, w_i)$, so $T^*I$ is a complex structure on
$T^*\bb C^n$ and $(z^1,\cdots, z^n, w_1,\cdots, w_n)$ are the
standard complex coordinates on $T^*\bb C^n$. Let $z=(z^1, \cdots,
z^n)^T$ and $w=(w_1, \cdots, w_n)^T$, then
\begin{eqnarray}
\omega &=& \mr{Re}\left( d\bar z^k\wedge d w_k\right)\cr &=& {1\over
2}(dz^\dag, dw^\dag)\left(\begin{matrix}0 & 1\cr -1 &
0\end{matrix}\right)\wedge\left(\begin{matrix}dz\cr
dw\end{matrix}\right)\cr &=& {1\over 2i}(du^\dag, d
v^\dag)\left(\begin{matrix}1 & 0\cr 0 &
-1\end{matrix}\right)\wedge\left(\begin{matrix}du\cr d
v\end{matrix}\right),
\end{eqnarray} where $u={z+iw\over \sqrt 2}$ and $v={z-iw\over \sqrt 2}$. Therefore, $\omega$ is always invariant under a complex
linear transformation $A$ on $T^*\bb C^n=\bb C^n\oplus \bb C^n$
satisfying the following condition:
$$
A^\dag \left(\begin{matrix}1 & 0\cr 0 & -1\end{matrix}\right) A=
\left(\begin{matrix}1 & 0\cr 0 & -1\end{matrix}\right).
$$
The collection of all such $A$'s form the group $\mr{U}(n,n)$ --- a
subgroup of $\mr{Sp}(4n, \bb R)$.  Note that $\mr{U}(1)$ acts on
$\bb C^n$ from right, and the induced action on $T^*\bb C^n$ can be
identified with the diagonal subgroup of
$\underbrace{\mr{U}(1)\times \cdots\times \mr{U}(1)}_{2n}$, so it
commutes with the action of $\mr{U}(n,n)$ on $T^*\bb C^n$. It is now
clear that Eq. (\ref{corres}) is really the decomposition in Eqs.
(4.1) and (4.2) of Ref. \cite{Howe} with the dual pair being
$(\mr{U}(1), \mr{U}(n,n))$ in $\mr{Sp}(4n, \bb R)$; consequently,
$\tilde{\ms H}(\sigma)$ is a unitary highest weight module of
$\widetilde{\mr U}(n,n)$. In view of the fact that $\tau$ in Eq.
(\ref{taumap}) is an isometry, by pulling back the action of
$\widetilde {\mr{U}}(n, n)$ on $\tilde {\ms H}(\sigma)$ via $\tau$,
we get the action of $\widetilde {\mr{U}}(n, n)$ on ${\ms
H}(\sigma)$. Part 5) is then proved.

\vskip 5pt
Since
\begin{eqnarray}
\omega = {1\over 2i}(du^\dag, d v^T)\left(\begin{matrix}1 & 0\cr 0 &
1\end{matrix}\right)\wedge\left(\begin{matrix}du\cr d\bar
v\end{matrix}\right),
\end{eqnarray} we know that $\omega$ is also invariant under the action
of $\mr{U}(2n)$ on $T^*\bb C^n=\bb C^n\oplus \overline{\bb C^n}$.
This $\mr{U}(2n)$, being the unitary group which leaves
$$|u|^2+|\bar v|^2=|z|^2+|w|^2$$ invariant, is a maximal compact subgroup of
$\mr{Sp}(4n, \bb R)$. It is not hard to see that
\begin{eqnarray}\label{maxcpt}\mr{U}(2n)\cap
\mr{U}(n,n)=\mr{U}(n)\times\mr{U}(n)
\end{eqnarray} with following identification:
\begin{eqnarray}\label{reverse}
\mr{U}(2n)\ni\left(\begin{matrix}A & 0\cr 0 & \bar
B\end{matrix}\right)\longleftrightarrow \left(\begin{matrix}A & 0\cr
0 & B\end{matrix}\right)\in \mr{U}(n,n).
\end{eqnarray}
The natural left action of $\mr{U}(n)$ on $\bb C^n$ induces a left
action on $T^*\bb C^n$. This induced action coincides with the
restriction to the diagonal subgroup of the above left action of
$\mr{U}(n)\times \mr{U}(n)$ on $T^*\bb C^n$. Therefore, the above
unitary action of $\widetilde {\mr U}(n, n)$ on the Hilbert space
${\ms H}(\sigma)$ of bound states, extends the manifest unitary
action of $\mr{U}(n)$
--- the diagonal of ${\mr U}(n)\tilde\times {\mr U}(n)$.

Since ${\mathcal H}_k$ is invariant under the action of $\tilde {\mr
U}(2n)$, in view of Eqs. (\ref{maxcpt}) and (\ref{corres}), we
conclude that the Hilbert space isomorphism
\begin{eqnarray}\label{identify}
\iota:\quad \bigoplus_{2I+|\bar\sigma|=k}\sigma^*\otimes\tilde{\ms
H}_I(\sigma)\to {\mathcal H}_k
\end{eqnarray} is an isomorphism of
$\mr{U}(n)\tilde\times\mr{U}(n)$-modules, so $\tilde{\ms
H}_I(\sigma)$ (hence ${\ms H}_I(\sigma)$) is a
$\mr{U}(n)\tilde\times\mr{U}(n)$-module. Since Eq. (\ref{energyE})
is a decomposition of $\mr{U}(n)$-modules, assuming that $\bar
\sigma\ge 0$, by using the Littlewood-Richardson rule, there is a
half integer $\kappa$ such that
\begin{eqnarray}{\ms H}_I(\sigma)\cong \overline{{\mathcal
R}_I^\kappa}\otimes{\mathcal R}_{I+|\bar\sigma|}^\kappa
\end{eqnarray} as representations of $\mr
U(n)\tilde\times \mr U(n)$. Therefore, in view of Eq.
(\ref{reverse}), the isomorphism in Eq. (\ref{identify}) and some
simple facts on harmonic isolators, we conclude that
$$
2I+|\bar\sigma|+n=2I+|\bar\sigma|+2n\kappa,
$$ so we must have $\kappa=1/2$. This proves part 3), and
consequently part 2) when $\bar\sigma\ge 0$. By the similar
arguments, the case when $\bar\sigma\le 0$ can be proved, too.

\appendix

\section{A formulae concerting the Laplace operators
}\label{appendix} Let $G$ be a compact connected reductive Lie
group, $X$, $B$ be manifolds, and $\pi$: $X\to B$ be a principal
$G$-bundle. For any $x\in X$, by linearizing $\pi$ at $x$ and map
\begin{eqnarray}
G &\to& X\cr g &\mapsto & x\cdot g^{-1}\nonumber
\end{eqnarray} at $e_G$, we arrive at short exact sequence
\begin{eqnarray}\label{exactseq}
0\to \frk{g}\to T_xX\to T_{\pi(x)}B\to 0.
\end{eqnarray}

We assume that $X$ is equipped with a Riemannian metric $g$ such
that $G$ becomes an isometric group of $(X, g)$.  This assumption,
via the orthogonal projection, yields an equivariant splitting of
(\ref{exactseq}) (i.e., a connection $\mathcal A$ on the principal
bundle) and consequently a quotient Riemannian metric on $B$.

Let $\sigma$ be a unitary irreducible representation of $G$, $V$ be
its underlying representation space, and $\xi$ be the corresponding
hermitian vector bundle. We use $\Delta$ to denote the
(non-negative) Laplace operator on $V$-valued smooth functions on
$X$ and $\Delta_{\mathcal A}$ to denote the (non-negative) Laplace
operator on smooth sections of $\xi$. Note that if $f$ is a smooth
section of $\xi$, then $f$ is actually a smooth map from $X$ to $V$
such that $f(x\cdot g^{-1})=\sigma(g)\cdot f(x)$ for any $(x, g)\in
X\times G$.

The following lemma can be proved by using the argument in appendix
A.1 of Ref. \cite{meng03}:
\begin{Lem}\label{LemAp} With the notations and background introduced above, we
have
$$
\Delta f= \Delta_{\mathcal A}f+c_2[G]|_\sigma f
$$ where $c_2[G]|_\sigma$ is the value of the Casimir operator of $G$ at
representation $\sigma$.

\end{Lem}
For example, we can let $X=G'$ be a compact Lie group and $G$ a
closed Lie subgroup of $G'$. Let $\xi$ be the vector bundle over
$B=G'/G$ associated with irreducible rep. $\sigma$ of $G$. Since
$\Delta=c_2[G']$, this lemma says that $ \Delta_{\mathcal
A}=c_2[G']-c_2[G]|_\sigma$ --- an identity proved in appendix A.1 of
Ref. \cite{meng03}.

\end{document}